\documentclass[12pt]{article}
\usepackage[margin=1in]{geometry} 

\usepackage[utf8]{inputenc}

\usepackage{todonotes}
\usepackage{xr-hyper,refcount}
\usepackage{graphicx}
\usepackage[colorlinks, citecolor=black, urlcolor=black, linkcolor=black]{hyperref}       
\usepackage{url}            
\usepackage{booktabs}       
\usepackage{amsfonts}       
\usepackage{nicefrac}       
\usepackage{microtype}      
\usepackage{amsmath,amsthm,amssymb,bbm}

\usepackage{libertine}
\usepackage{libertinust1math}
\usepackage[T1]{fontenc}
\usepackage{titlesec}
\titleformat{\section}
  {\normalfont\fontsize{16}{16}\bfseries}{\thesection}{1em}{}
  \titleformat{\subsection}
  {\normalfont\fontsize{14}{14}\bfseries}{\thesubsection}{1em}{}

\usepackage{enumitem}
\setlist[itemize]{noitemsep, nolistsep}

	\usepackage[sort&compress]{cleveref}
  \Crefname{chapter}{section}{sections}
	\Crefname{chapter}{Section}{Sections}
	\Crefname{figure}{Figure}{Figures}

\title{Diversity in Sociotechnical Machine Learning Systems}
\author{Sina Fazelpour\\ \small{Northeastern University}\\ \small{s.fazel-pour@northeastern.edu}  \and Maria De-Arteaga \\ \small{University of Texas at Austin}\\ \small{dearteaga@mccombs.utexas.edu}
}
\date{}

\usepackage[numbers]{natbib}
\begin{document}

\maketitle

\begin{abstract}
  
  There has been a surge of recent interest in sociocultural diversity in machine learning (ML) research, with researchers (i) examining the benefits of diversity as an organizational solution for alleviating problems with algorithmic bias, and (ii) proposing measures and methods for implementing diversity as a design desideratum in the construction of predictive algorithms. 
  Currently, however, there is a gap between discussions of measures and benefits of diversity in ML, on the one hand, and the broader research on the underlying concepts of diversity and the precise mechanisms of its functional benefits, on the other. 
  This gap is problematic because diversity is not a monolithic concept. Rather, different concepts of diversity are based on distinct rationales that should inform how we measure diversity in a given context. Similarly, the lack of specificity about the precise mechanisms underpinning diversity's potential benefits can result in uninformative generalities, invalid experimental designs, and illicit interpretations of findings.
  In this work, we draw on research in philosophy, psychology, and social and organizational sciences to make three contributions: First, we introduce a taxonomy of different diversity concepts from philosophy of science, and explicate the distinct epistemic and political rationales underlying these concepts. Second, we provide an overview of mechanisms by which diversity can benefit group performance. Third, we situate these taxonomies—of concepts and mechanisms—in the lifecycle of sociotechnical ML systems and make a case for their usefulness in fair and accountable ML. We do so by illustrating how they clarify the discourse around diversity in the context of ML systems, promote the formulation of more precise research questions about diversity's impact, and provide conceptual tools to further advance research and practice.
\end{abstract}


\maketitle
\pagestyle{plain} 

\section{Introduction} \label{sec:intro}
Sociocultural diversity is a key value in democratic societies both
for reasons of justice, fairness, and legitimacy and because of its
ramifications for group performance. As a result, researchers in humanities and social sciences have worked to understand diversity’s varied meanings, develop appropriate measures for quantifying diversity (in some sense), and specify pathways by which diversity can be functionally beneficial to groups such as juries, design teams, and scientific communities.\footnote{A terminological note: in what follows we frequently use the term ``epistemic''. In philosophy jargon, the term refers to knowledge-related phenomena, broadly construed. Accordingly, \textit{epistemic} activities include information gathering, learning, problem-solving, deliberation, decision-making, and so on. By epistemic agents and communities, we mean agents and communities engaged in such activities. And by epistemic performance we mean the performance on these activities (according to some criteria).} Recently, there has also been a surge of interest in sociocultural diversity in machine learning (ML) research. This burgeoning literature on diversity\footnote{Throughout when we use diversity, we mean diversity along sociocultural lines. We explicitly qualify the term in contexts where there is a possibility of confusion, for example, between sociocultural and \textit{task-relevant} or \textit{cognitive} diversity.} in ML systems can be roughly divided into two general lines. 
One set of diversity-related considerations pertains to the composition and dynamics of teams and groups whose judgments shape the construction of ML systems---e.g., those engaged in problem formulation, data generation, and development. From this perspective, there have been claims about the potential benefits of diversifying these teams as an organizational solution to alleviate biases in ML systems \citep{jobin2019global}, with more recent efforts aiming to empirically test these purported benefits \citep{duan2020does, cowgill2020biased}.
A second set of issues concerns the composition of items at different stages of the data-processing pipeline, especially in relation to who or what gets represented therein and how it affects individuals and communities that are impacted by the deployment of ML systems.
In the context of curating input data, for instance, the lack of geographic diversity in benchmark image datasets has been linked to amerocentric and eurocentric algorithmic biases, such as higher misclassification rates for bridegroom images from Pakistan than from the US
~\citep{shankar2017no}. Similarly, recent works have highlighted the importance of diversity in relation to the output of subset selection across tasks ranging from image search and content recommendation to matchmaking and automated recruiting~\citep{drosou2017diversity}. 
To incorporate these considerations into the design of ML products, researchers have proposed various measures for quantifying diversity, developed methods for satisfying these measures, and examined the interaction between diversity and other design desiderata such as predictive performance and fairness~\citep{drosou2017diversity, celis2016fair, mitchell2020diversity}.

Currently, however, the discourse around diversity in ML systems is hindered by a lack of clarity regarding the underlying concepts of diversity and the precise pathways by which diversity can benefit group performance \citep{jobin2019global}. This is problematic because diversity is an ambiguous term that admits various, potentially conflicting, conceptions. These diversity concepts differ substantially in their motivations, meanings, and appropriate operationalization. Without explicitly grounding proposed diversity measures in appropriate underlying concepts, we thus risk a mismatch between professed diversity-related aims and the operationalization and implementation of those aims. This disconnect can not only lead to the use of contextually inappropriate measures, it can also result in oversimplified claims about how considerations of diversity interact with other desiderata such as predictive performance and fairness. What is more, whether and when diversity (in some sense) can improve the performance of a collective requires attending to the precise pathways of diversity's influence, the specific factors that moderate this impact, and the broader enabling conditions that support and sustain these potential positive consequences. The lack of specificity about the precise mechanisms underpinning diversity’s potential benefits can thus result in espousing uninformative or easily falsifiable generalities, invalid experimental designs, and illicit inferences from findings.

In this paper, we bridge the gap that currently exists between discussions of measures and beneficial consequences of diversity in ML, on the one hand, and the humanities and social sciences research on concepts and consequences of diversity, on the other. We begin by articulating current and historical thinking about the concepts and consequences of sociocultural diversity in feminist philosophy, social psychology, organizational and network sciences. Drawing on this literature, in \Cref{sec:concepts}, we present a critical survey of
attempts to provide a classification of different conceptions of sociocultural diversity. In distinguishing between different diversity concepts, we call attention to their distinct and potentially conflicting epistemic, ethical, and political rationales, and present popular families of diversity measures in use for quantifying them. In \Cref{sec:benefits}, we discuss mechanisms through which diversity can potentially improve the performance of teams and collectives, with a focus on two types of pathways. For each, we describe the logic of the pathway and the evidence for potential benefits, and highlight the key effect modifiers that moderate diversity’s beneficial effects along that pathway. We end the section by emphasizing institutional and societal enabling conditions that are critical for realizing diversity's potential benefits in practice. In \Cref{sec:ml-cycle}, we situate this understanding in and draw its implications for the discussions of diversity in ML. In mapping the various types of diversity-related considerations that arise throughout the ML lifecycle, we draw attention to the significance of achieving clarity about the conceptual underpinnings of diversity measures and the precise mechanisms mediating diversity's potential benefits. We show how this can enrich the design and evaluation of ML systems by informing diversity-related hypotheses formulation, experimental design, and interpretations of findings.

Before we proceed two qualifications are in order. First, diversity is a topic of research in a variety of disciplines. The discussions below mainly draw from research on diversity in philosophy, psychology, sociology, and organizational science. This focus should \textit{not} be taken as implying that research on diversity in other disciplines (e.g., library sciences or ecology) are unimportant or have little to offer to discussions of diversity in ML. Rather, it is a simple admission of the fact that diversity research is vast, and our expertise is limited. Second, diversity is valued on many different moral, political, social, and epistemic grounds. Discussions of the potential epistemic benefits of team diversity should \textit{not} be confused with the ``business case'' for diversity---the idea that institutions should value diversity because, and insofar as, it can improve their performance. The ``business case" takes place against a particularly troubling moral background: it ignores other---and as we suggest in \Cref{sec:benefits} more fundamental---ethical and political grounds for valuing diversity. What is more, it takes as default the \textit{lack} of diversity, and puts the burden on individuals from marginalized communities to justify their presence by demonstrating how their inclusion results in performance benefits. The moral background against which discussions of the epistemic benefits of diversity take place thus matters. As we will discuss in detail in \Cref{sec:concepts}, there is a long line of research in feminist and standpoint epistemologies that has not only emphasized diversity's epistemic benefits, but also highlighted the ways in which they work in concert with ethical and political considerations of social justice, anti-oppression, and participation. Importantly, beneficial epistemic impacts of diversity are broader than---and may come apart from---those relevant to the ``business case". As we note in \Cref{sec:benefits}, for example, one way in which diversity can enhance group performance is by promoting mechanisms that support critical and contestable collective inquiry. These tend to increase reliability by \textit{delaying} convergence and may very well be in conflict with business aims involving speed and efficiency.\footnote{More generally, assuming that the epistemic case simply maps onto the business case is analogous to supposing that community-level norms and structures that are conducive to good research and innovation are things that all and only businesses care about.} 
\section{The varied concepts of diversity} \label{sec:concepts}
Diversity implies some sort of difference. Divergent understandings of what this difference consists in lead to varied meanings and perceptions of diversity. 
Perhaps the most obvious way our assessments of a collective's diversity can vary depends on what we take to be the set of attributes along which its members exhibit relevant differences. So, the same group can be seen as more or less diverse depending on the attribute (e.g., gender, ethnicity, socioeconomic status) used for characterizing its members.\footnote{Of course, individuals differ from one another across various dimensions. One strand of diversity research has thus sought to propose classification schemes for distinguishing between types or categories of attributes that might matter to the study of diversity. Such classification schemes often take the form of two-category classifications \citep{roberson2019diversity}. So, researchers have drawn distinctions between, among others, surface \textit{vs.} deep diversity \citep{harrison1998beyond}, identity \textit{vs.} functional (or cognitive) diversity \citep{hong2004groups}, and social category \textit{vs.} informational diversity~\citep{phillips2017real}. While this line of research has often focused on single attributes (e.g., race or gender), a complementary line of work on ``demographic faultlines''~\citep{lau2005interactions} has examined the alignments and interactions among multiple attributes (e.g., race \textit{and} gender).} 
As noted by a number of researchers, however, even abstracting from questions of relevant attributes, the notion of diversity admits multiple understandings that can differ significantly in their underlying rationales and appropriate operationalizations \citep{harrison2007diversity, page2010diversity, steel2018multiple}. In particular, \citet{steel2018multiple} distinguish between \textit{contexts}, \textit{concepts}, and \textit{measures} of diversity. The \textit{context of diversity} refers to a background situation that suggests, among other things, a set of relevant attributes. A \textit{concept of diversity}, in contrast, consists of ``an understanding of what constitutes diversity, abstracted from questions of which attributes are relevant to diversity in a specific context'' \cite[p.726]{steel2018multiple}. \textit{Measures of diversity} are mathematical operationalizations for quantifying the extent of diversity according to some specific concept. Accordingly, the concept-measures relation is often one-many, in that the construct validity of a given measure depends on the extent to which it tracks a specific underlying concept in the appropriate way. 

While there are different ways of drawing distinctions between diversity concepts, in what follows we primarily build on the classification developed by \citet{steel2018multiple}, which is purpose-built for understanding \textit{sociocultural} diversity and its potential epistemic benefits for groups. Throughout, we also incorporate discussions from other works. 
We start with general families of diversity concepts, discuss prevalent types of concepts within each family, and highlight the motivations behind these concepts as well as some popular ways of operationalizing them.

\subsection{Within-group family of concepts}
Suppose there is a focal group of individuals, $G$, whose diversity is a matter of question. $G$ might be a group of stakeholders deliberating about how to formulate a social problem, or a group of crowdsource workers engaged in labeling some type of content. 
For simplicity, let us assume that, given the context, there is an attribute, $A \in \{a_1, a_2, ..., a_n\}$, which is seen as relevant for assessing $G$'s diversity. $A$ might refer to gender, race, socioeconomic status, and more. According to \textit{within-group family} of diversity concepts, assessments of $G$'s diversity can proceed by mainly focusing on the properties of $G$ itself. To make this concrete, we examine \textit{egalitarian diversity concepts}, which are, arguably, the most widely used type in the within-group family.\footnote{We focus on egalitarian concepts because of their prevalence in research and application as well as their relevance to our discussion. But the family of within-group concepts includes other types (e.g., separation and disparity introduced by \citet{harrison2007diversity}).}

\paragraph{Egalitarian concepts.}
Broadly speaking, egalitarian concepts are concerned with \textit{balanced representation} of sub-groups within a focal group. 
Typically this concern pertains to to the presence as well as the proportion of each attribute category $a_i \in A$ present in $G$.\footnote{\citet{harrison2007diversity} refer to this concept as \textit{variety}.} In general, diversity is maximal, in the egalitarian sense, when the distribution in $G$ is uniform over $A$. Most egalitarian concepts thus tend to be \textit{symmetric}, in the sense that changing the referent of two category attributes, $a_i$ and $a_j$, does not affect the diversity of $G$. In practice, there are various operationalizations of egalitarian concepts, particularly in terms of different specifications of generalized entropy functions \citep{page2010diversity}. Different parameterizations of these functions leads to measures---such as Blau's index \citep{blau1977inequality} in sociology, Simpson's index \cite{simpson1949measurement} in ecology, and the effective number of parties \cite{laakso1979effective} in political science---that may assess departures from the uniform distribution in different ways.  

Egalitarian concepts can be enriched in various ways. For example, in addition to considering the presence and proportion of categories, one might assign a weight to each attribute category. This might be desirable, for instance, because of the contextual relevance of the extent of (dis)similarity between various attribute categories (e.g., different ethnicities). One way of accommodating this desideratum within egalitarian concepts is to assign a weighting to categories on the basis of some well-defined similarity function \cite{stirling2007general,page2010diversity,steel2018multiple}.\footnote{Other authors consider the notion of \textit{diversity as difference} to be a separate concept, distinct from egalitarian notions \citep{weitzman1992diversity,harrison2007diversity}. Arguably, whether (the extent of) similarities should constitute an entirely different concept or a weighting on egalitarian concepts depends on the aims of researchers.} 

What are the rationales for conceiving diversity in this way? As their name suggests, a typical political and ethical rationale for egalitarian conceptions is the ideal of equal participation. In feminist philosophy of science, moreover, this ethical rationale is often tied to proposed epistemic benefits \citep{longino1990science, solomon2007social, grasswick2018feminist-social-epistemology}. A core insight of this literature is the recognition of the \textit{situated nature of knowledge}---the idea that individual agents are constrained by their situation in ways that induce relevant differences in the set of hypotheses that they might pursue, their knowledge base, their task-relevant expertise, and more. Similar points are also expressed in strands of diversity research from sociology and organizational research \citep{page2017diversity, cox1991managing}. Viewed in this way, therefore, egalitarian conceptions of diversity are valuable because the variety of perspectives and approaches promotes a more comprehensive pursuit of alternative hypotheses \citep{solomon2007social}, reduces the risks of inadvertently accepting unjustified assumptions \citep{longino1990science}, and enables a fruitful way for groups to leverage the benefit of exploration \citep{hong2004groups}.

\subsection{Comparative family of concepts}
The assessments of diversity via within-group concepts almost solely focus on the properties of a focal group. 
In many circumstances, however, the narrow focus of egalitarian conceptions fail to capture broader social features that motivate our analyses of sociocultural diversity in the first place \citep{rushton2008note}. As a simple example, suppose $A \in \{a_1, a_2, a_3, a_4\}$ refers to ethnicity in a social context where individuals from $a_1$ constitute $40\%$ of the general population, with the rest equally consisting of $a_2$, $a_3$ and $a_4$. Viewed through the lens of egalitarian diversity concepts, a balanced group, $G_1$, consisting of $25\%$ from each category, is maximally diverse. Furthermore, for symmetric versions, the extent of (egalitarian) diversity remains unperturbed when we shift our attention from $G_2 = \{0.6, 0.2, 0.2, 0\}$ to $G_3 = \{0, 0.2, 0.2, 0.6\}$. The core idea behind \textit{comparative diversity concepts} is that, in many circumstances, our understanding of diversity needs to be sensitive to broader societal information that could render $G_2$ and $G_3$ significantly different. In assessing a focal group's diversity, \textit{comparative} concepts thus require additional contextual information. 
Below, we examine two types of diversity concepts that aim to incorporate this idea, albeit in different ways.

\paragraph{Representative concepts.}
Given a prior designation of a reference population $F$, a focal group $G$ is diverse, in the \textit{representative} sense, to the extent that its distribution over $A$ \textit{matches} that of $F$ \citep{jacklin1978representative, steel2018multiple}.\footnote{For a broader discussion of the manifold meanings of representativeness, see \citet{chasalow2021representativeness}.} The choice of $F$ is often value-driven. In the context of a group subject to admission decisions, for example, $F$ might be taken as the general population, the population of university graduates in relevant fields, or previous applicant pools. Accordingly, appraisals of $G$'s diversity can vary depending on what is taken to be the relevant reference population. Given $G$ and $F$, different distance or proportionality measures (e.g., demographic parity) can be used for constructing quantitative measures of representative diversity. While the latter measures are familiar in the context of fair ML in relation to discussions of disparate impact, as we discuss in Section \ref{sec:discussion}, the surface similarity in mathematical formulation should not obscure fundamental differences in what the measures actually mean. 

As before, claims about the relevance of the representative conception can be motivated on different grounds. The obvious ethical and political rationale emerges from the association of this conception with political ideals of \textit{representative democracy}. In certain contexts, the representative concept might also be desirable for epistemic reasons. 
When selecting a group of experts on the basis of their views on an issue about which there is robust evidence, for example, it might be argued that the relevant sense of diversity is representative, as opposed to egalitarian, insofar as the latter can give rise to an \textit{overrepresentation} of unfounded outlier views \citep{steel2018multiple, oreskes2011merchants, voakes1996diversity}. 

\paragraph{Normic concepts.} 
While representative concepts take a broader perspective on diversity, contextually critical considerations of social dynamics still remain absent from this understanding. Consider again, groups $G_2 = \{0.6, 0.2, 0.2, 0\}$ and $G_3 = \{0, 0.2, 0.2, 0.6\}$ in a population with a $\{0.4, 0.2, 0.2, 0.2\}$ proportions. While $G_2$ and $G_3$ remain indistinguishable from an egalitarian perspective (absent any weightings), $G_3$ might be seen as more diverse of the two in the representative sense. 
Suppose, however, that the sub-group characterized by category $a_1$ is not only a relative numerical majority (with $40\%$), but also a dominant group, which is historically over-represented in positions of power. Suppose that $a_4$ in contrast refers to a marginalized group, whose members have historically been excluded from socially important decision-making processes due to unjust and oppressive means. Given this setting, in certain contexts, we might view $G_3$ as \textit{more diverse} than $G_2$. 

Capturing these sociopolitical and historical dimensions of diversity is at the heart of \textit{normic diversity concepts}. One way of proceeding is by characterizing diversity in comparison with a \textit{non-diverse category} in the reference population \citep{steel2018multiple}. In the example above, a focal group can be seen as diverse to the extent that its members \textit{diverge} from the dominant, non-diverse category, $a_1$. Alternately, one can assess the diversity of a focal group in terms of the extent to which it \textit{matches} an \textit{ideal} distribution over other attribute categories (e.g., a distribution that is skewed towards the historically marginalized group $a_4$) (e.g., \citep{harding2015objectivity}).
While there are many potential ways of formalizing this idea (e.g., in terms of divergence measures), compared to other conceptions of diversity, measures of normic concepts remain rare. Instead, the motivation behind normic concepts is often incorporated in terms of weightings on attribute categories in within-group concepts, resulting in hybrid conceptions. A rare exception is \citet{mitchell2020diversity}, who introduce diversity metrics for subset selection corresponding to a normic concept, which we discuss in more detail in \Cref{sec:ml-cycle}. 

From ethical and political perspectives, the motivation behind normic concepts is grounded in ideals of social justice, inclusion, and anti-domination. In addition, a long line of research in standpoint theory has highlighted an epistemic rationale for these conceptions~\citep{harding2004feminist,collins2002black,wylie2003standpoint}. Besides recognizing the socially situated nature of knowledge discussed above, standpoint theorists often endorse an \textit{epistemic advantage thesis}, according to which ``[s]ome standpoints, specifically the standpoints of marginalized or oppressed groups, are epistemically advantaged (at least in some contexts)''~\citep[p. 783]{intemann201025}. 
The core idea is that those occupying the standpoints of marginalized groups can develop a more accurate and thorough understanding of institutional knowledge production and decision-making processes~\citep{harding2004feminist}. In contrast, as beneficiaries of the status quo, dominant standpoints can result in significant blind spots or else distorted understandings of these processes~\citep{grasswick2018feminist-social-epistemology}. To be sure, as many standpoint theorists have emphasized (e.g.,~\citep{wylie2003standpoint}), the epistemic advantage thesis is not an a priori claim. Rather, whether, how, and to what extent some marginalized standpoints afford epistemic advantages is a context-dependent question, open to empirical investigation~\citep{grasswick2018feminist-social-epistemology}. The thesis has found empirical support in various aspects of biomedical and social scientific research~\citep{intemann201025, wylie2017knowers, komaromy1996role}. 

It is worth noting that the notion of a \textit{standpoint} in standpoint theory should not be equated with that of a \textit{perspective} that one occupies simply by virtue of belonging to a demographic group (e.g., gender or race). The distinction is driven by concerns that such an equation results in misleading essentialist views on cognitive differences between demographic groups. Instead, as standpoint theorists emphasize, adopting the standpoint of marginalized groups, while made more likely by group membership, requires active engagement with the relevant societal issues~\citep{grasswick2018feminist-social-epistemology}. Finally, notice that unlike the diversity concepts considered so far, where diversity is strictly speaking a characteristic of groups, one could properly refer to an \textit{individual} (e.g., a job applicant, a scientist, etc.) as diverse in the normic sense. Importantly, however, the above rationales behind normic concepts prevent their use in a tokenizing way.

\subsection{Combine, but don't conflate}
Even abstracting away from the choice of relevant attributes, then, the notion of diversity can have varied meanings.
Depending on context, it might be perfectly sensible to combine these concepts (e.g., constructing hybrid concepts by drawing on normic considerations to inform the weighting on egalitarian concepts) \citep{steel2018multiple}. While such combinations can be permissible, it is critical not to \textit{conflate} different concepts. This is because different diversity concepts (and the measures generated to formalize them) are based on distinct rationales, and make fundamentally different assumptions about which aspects of the social environment are relevant for, or can be abstracted away from, evaluations of diversity. Aside from the conceptual confusion it can give rise to, conflating different concepts can also lead to illicit inferences from empirical research \citep{steel2018multiple}. 
Attending to these considerations is particularly important, since, as noted by some diversity researchers (e.g., \citep{page2010diversity}), the current widespread use of certain egalitarian measures (e.g., Blau's index) is mainly driven by convention, instead of intention. These considerations are also critical when building on recent calls for increasing diversity in ML as a way of improving design and combating bias. 
With this, we now turn to works examining ways in which diversity can benefit groups. 
\section{Potential epistemic benefits of diversity} \label{sec:benefits}
As discussed above, in addition to referring to distinct ethical and political ideals, researchers often motivate different concepts of diversity by highlighting how increasing diversity \textit{can} improve the \textit{epistemic} performance of groups. This emphasis on the word ``can'', and the implied distinction between \textit{potential} and \textit{actual} benefits, is key~\citep{phillips2017real}:~diversity's influence on group performance is mediated through various cognitive, communicative, and affective pathways. The mode of diversity's influence (positive, negligible, or negative) along each pathway critically depends on the type of outcomes used for measuring performance, pathway-specific effect modifiers, and broader contextual features. In light of these multiplicities---of performance measures, pathways and effect modifiers---anticipating the aggregate consequences of diversity in context is challenging. 
When exploring diversity's potential epistemic benefits, it is thus necessary to be explicit about the specific \textit{mechanism} under investigation~\citep{roberson2019diversity}. 

We start this section by distinguishing between two types of pathways---\textit{cognitive} and \textit{information elaboration}---through which sociocultural diversity can improve the epistemic performance of groups. In each case, we describe the core idea, outline examples of supporting evidence, and highlight some key effect modifiers pertinent to these pathways. We finish this section by discussing the seemingly conflicting findings of various meta-analytical studies of sociocultural diversity and the significance of context.

\subsection{Cognitive pathways}
The core idea behind cognitive pathways is that diverse groups can leverage the variety of the cognitive repertoires of their members---their varied knowledge, perspectives, skills, etc.---to more effectively deal with demanding tasks. This idea underpins the epistemic rationale for many of the diversity concepts discussed in Section \ref{sec:concepts}. 

This general idea can be analyzed in terms of a series of interrelated steps. First, the informational and processing demands of certain tasks mean that while some agents may be more likely to succeed at the task, there is often no individual whose set of relevant abilities guarantees optimal performance across all relevant variations of the task. Such tasks are thus better dealt with by groups rather than any single agent. Individuals in cognitively homogeneous groups, however, tend to exhibit significant overlap in their relevant strengths and limitations~\citep{bang2017making}. \textit{Cognitive} diversity can be beneficial, particularly because of the complementarity between individual strengths \citep{page2017diversity}. 
Notably, insofar as many complex tasks require striking a balance between exploiting current best methods and exploring alternatives, cognitive diversity can be beneficial, even when, on average, members of the heterogeneous group exhibit lower capabilities, according to some measure, than those of the homogeneous one~\citep{hong2004groups, grim2019diversity}.\footnote{Notice the similarity between these ideas and ideas underpinning some ensemble learning approaches in ML.} \textit{Sociocultural} diversity can enhance the epistemic performance of groups, via these cognitive pathways, insofar as it can induce cognitive diversity---that is, to the extent that it influences or correlates with the variety in task-relevant knowledge and skills~\citep{page2017diversity}.

The potential benefits of sociocultural diversity via cognitive pathways have been demonstrated by different lines of evidence. Empirical support has been found in relation to different tasks and groups~\citep{roberson2019diversity}. In organizational contexts, for example, Milliken and Martins review studies demonstrating the beneficial effects of sociocultural diversity in relation to cognitively relevant outcomes such as the number and quality of alternative ideas considered by groups~\citep{milliken1996searching}. Using citation counts as a proxy for research quality,~\citet{campbell2013gender} found evidence for the positive impact of gender diversity on the quality of work produced by scientific teams.\footnote{Note that most of these studies employ a within-group, and especially egalitarian, notions of diversity.} 

In addition to these empirical studies, researchers in cognitive science~\citep{bang2017making}, sociology~\citep{page2017diversity}, and philosophy~\citep{muldoon2013diversity} have turned to formal and simulation tools to investigate the impacts of diversity along cognitive pathways. These studies typically center around frameworks (e.g., NK landscapes, multi-armed bandits) that represent tasks that demand exploitation-exploration trade-offs. While these studies often abstract away from the relation between sociocultural and cognitive diversity, they nonetheless offer valuable insights about the conditions---e.g., task difficulty, team composition, network structure---under which cognitive diversity can benefit group performance in tasks such as learning and problem-solving~\citep{zollman2007communication, hong2004groups, lazer2007network, weisberg2009epistemic}.  

In addition to pointing towards the potential epistemic benefits of sociocultural diversity, these lines of research also identify key contextual factors that moderate diversity's benefits along cognitive pathways as well as the potential trade-offs involved (e.g., cognitive diversity vs. stability). As emphasized by both empirical~\citep{van2012defying} and simulation~\citep{hong2004groups} studies, for example, chief among these effect modifiers are \textit{task demands and structure}. When tasks are trivial to an agent (due to task simplicity or prior expertise), there is often no need to rely on others~\citep{hoppitt2013social,kendal2018social}.  
Similarly, highly regimented tasks consisting of a set of prescribed steps do not provide agents with the requisite opportunity to draw on their potentially distinct abilities~\citep{van2012defying}. All else equal, therefore, diversity is more likely to result in performance benefits in complex tasks and those that demand innovation and creativity.\footnote{Indeed, even with respect to these, the potential benefits of collaborative, as opposed to individual, work (whether in homogeneous or diverse groups) might depend on the particular cognitive abilities called upon by the task as well as factors influencing communication~\citep{gyory2019you}.} 
In Section \ref{sec:ml-cycle}, we will discuss the importance of these considerations for studies that seek to explore the potential epistemic benefits of diversity in the context of ML systems.

\subsection{Information elaboration pathways}
The performance of any task by a group requires the elicitation, examination, and ultimately integration of information that is distributed among individual members. In diversity research, these processes are collectively referred to as \textit{information elaboration}~\citep{homan2007bridging, kooij2008ethnic, steel2019information}. 
Importantly, even when cognitively diverse, socioculturally homogeneous groups are particularly susceptible to group influences that undermine efficient information elaboration. By reducing these detrimental effects of homogeneity, therefore, sociocultural diversity can epistemically benefit groups, even when not correlated with cognitive diversity in context~\citep{phillips2017real, steel2019information}.

This idea can be made precise by examining how specific aspects of the information elaboration process can be adversely affected by dynamics of socioculturally homogeneous groups. 
Consider how sociocultural homogeneity can prevent effective information elicitation and sharing. Here, individuals tend to take perceived similarity in social markers of identity as an indicator of similarity in underlying cognitive repertoire (assumptions, information, ways of thinking, ...)~\citep{phillips2017real}. 
Individuals in homogeneous groups may thus \textit{overestimate} the extent to which their peers possess or share the same information or views, resulting in omissions, whereby individuals fail to share key pieces of information with peers.\footnote{In this respect, also see the literature on groupthink~\citep{janis2008groupthink}.} 
In contrast, perceived differences along identity markers can lead to \textit{expected cognitive differences} in diverse groups, even if this is not in fact the case~\citep{phillips2017real, roberson2019diversity}. Such expected cognitive difference have been shown to reduce the likelihood of critical omissions in information sharing~\citep{phillips2006surface}. 

Information sharing in homogeneous groups is further adversely impacted by \textit{conformity} pressures. Driven by a desire to belong to a group or to avoid punishment, these conformity pressures often take the form of a \textit{normative expectation} that individuals ought to agree with the (imagined) group consensus \citep{deutsch1955study, cialdini2004social}. As a result, individuals in homogeneous groups are less likely to voice dissenting opinions. The detrimental impact of conformity on information sharing is further exacerbated by the fact that individuals tend to be less receptive of and particularly incensed by dissenting opinions of in-group peers \citep{phillips2017real}. By reducing these negative impacts of group-based conformity on the sharing and reception of dissenting views, sociocultural diversity has been shown to improve group performance \citep{phillips2009pain}. Importantly, as shown by Phillips et al. \citep{phillips2009pain} in this context, while sociocultural diversity improves group performance according to objective measures (e.g., correctness of solutions), it is also associated with negative impact when assessed using subjective measures (e.g., subjective levels of discomfort and task-related uncertainty). These findings are crucial when thinking about the choice of dependent outcomes used to measure diversity's impact as well as in relation to broader enabling conditions for actualizing diversity's potential benefits. We will return to these points below. 

Homogeneity can also adversely impact information processing. 
In many social settings, individuals appear to consider in-groups as more reliable and trustworthy sources of information about the world \citep{turner1989referent, turner1991social}. This mechanism of group-based trust can undermine the epistemic performance of homogeneous groups. In particular, the uncritical reliance on in-groups results in reduced vigilance in processing social information, hasty belief revisions in light of (potentially misleading) information, and ultimately premature consensus. 
Once more, empirical studies have shown that sociocultural diversity can counteract these detrimental influences on group performance by enhancing the critical assessment of incoming information~\citep{gaither2018mere, levine2014ethnic}.\footnote{There other additional pathways by which diversity can enhance information elaboration, and so performance. For example, a separate line of research highlights the importance of democratic turn-taking and interaction styles on group performance in information sharing---factors that are found to be positively correlated with gender diversity \citep{woolley2010evidence, bear2011role}.}

Similar to the case of cognitive pathways, these empirical investigations are complemented by results from simulation studies \citep{fazelpour2021diversity}. 
Together, these empirical and simulation studies not only demonstrate the epistemic benefits of diversity via information elaboration pathways, but also emphasize their relation to democratic ideals of contestation and critical uptake~\citep{fazelpour2021diversity,anderson2006epistemology}. 
Incorporating the insights from these works can be fruitful when thinking about the influence of diversity in the context of designing ML systems. This is particularly the case when there is interaction between different individuals (e.g., stakeholders, experts, crowdsource workers, or developers), and where individuals are either aware of others' identities or else might have prior expectations about what this identity might be (e.g., by assuming that they too are members of the same statistically prevalent group). 
In drawing from this literature, it is nonetheless important to attend to potential trade-offs (e.g., speed vs. reliability of convergence~\citep{fazelpour2021diversity}) that affect diversity's benefits via information elaboration pathways as well as unique communicative challenges faced by diverse groups~\citep{phillips2017real}.

\subsection{Diversity's overall impact and the significance of context}
We have highlighted some mechanisms by which diversity \textit{can} benefit group performance. To what extent do these potential benefits translate beyond the settings of experimental labs and simulations to everyday teams and communities? According to meta-analytical studies, the overall impact of sociocultural diversity, and to a lesser extent cognitive diversity, on task performance is highly contextual and tends to be mixed---ranging from positive to slightly negative \citep{van2012defying, horwitz2007effects, bell2011getting, bear2011role, post2015women}. 
Some of the reasons for these mixed findings have already been mentioned above. In Section \ref{sec:concepts}, for example, we emphasized the importance of employing appropriate concepts and measures of diversity, particularly since many conventional measures are insensitive to contextual differences between groups~\citep{steel2018multiple}. In Section 3.2, we mentioned the importance of the choice of dependent outcomes used to measure diversity's impact. For instance, sociocultural diversity's impact can be positive when measured using objective task-relevant outcomes (e.g., correctness and quality of solutions), but negative when probed via subjective outcomes (e.g., feelings of discomfort or uncertainty)~\citep{phillips2009pain,van2012defying}. 

In many experimental settings, the interaction between these objective and subjective outcomes might be minimal. But this is not necessarily so when we turn to longer term interactions in teams and communities embedded in contexts with entrenched power asymmetries~\citep{fazelpour2021diversity}. Negative subjective outcomes (e.g., feelings of discomfort) can affect objective group performance, for instance, by increasing the possibility of organizational gate-keeping by dominant groups, thus changing the very structure and composition of groups \citep{anicich2021structuring}. Moreover, in many real-world settings, such subjective outcomes are core determinants of how contributions from the members of different groups are evaluated. For example, through an analysis of three decades of PhD dissertations in the US, Hofstra et al. find that while students from demographically underrepresented groups were more likely than majority students to produce innovative works, ``their novel contributions are discounted and less likely to earn them academic positions''~\citep[p. 9284]{hofstra2020diversity}.

These issues provide an important occasion for examining broader institutional and societal settings wherein diverse groups can successfully function. For example, the positive impact of gender diversity is even more pronounced in inclusive settings with egalitarian power distributions \citep{post2015women} (see also \citep{bear2011role, joshi2009role}). More broadly, increased sociocultural diversity is more likely to improve group performance when group members are open to social differences; 
when groups are less homogeneous to begin with; when groups function in settings with more egalitarian (rather than hierarchical) power dynamics; and when individuals from marginalized demographics find support in an organizational leadership that values and promotes a culture of inclusion \citep{phillips2017real}. Absent such enabling conditions, ``diversifying'' teams by simply introducing individuals from underrepresented or marginalized backgrounds, and then expecting improvements to team performance, may amount to little more than setting up these individuals for failure~\citep{phillips2017real}.

Insofar as performance benefits are concerned, therefore, sociocultural diversity is not a panacea. Examining diversity's potential benefits requires specifying the relevant pathways and effect moderators affecting these pathways. What is more, diversity's positive impacts are highly contingent upon details of the task, the mode and duration of interactions, adopted performance measures, and broader contextual factors. It is worth belaboring these points, lest the value of diversity is tied to its potential epistemic benefits. Indeed, as the foregoing makes clear, the epistemic benefits of sociocultural diversity are more likely to emerge precisely in those contexts where diversity is already valued on other grounds.  
\section{Situating diversity in sociotechnical ML systems} \label{sec:ml-cycle}
In this section, we situate the discussion of concepts of diversity and mechanisms of its benefit in ML by mapping the various contexts in which sociocultural diversity can be relevant throughout the ML lifecycle. We emphasize that diversity-related considerations are ubiquitous in ML systems. While often neglected, this fact should not come as a surprise; these systems are sociotechnical systems, embedded in decision pipelines that are shaped by, or else implicate, groups and communities. 
Nonetheless, identifying, understanding, and implementing these diversity-related considerations hinges on being specific about concepts, rationales, pathways, and more. 

\begin{figure}
    \centering
    \includegraphics[width=\textwidth]{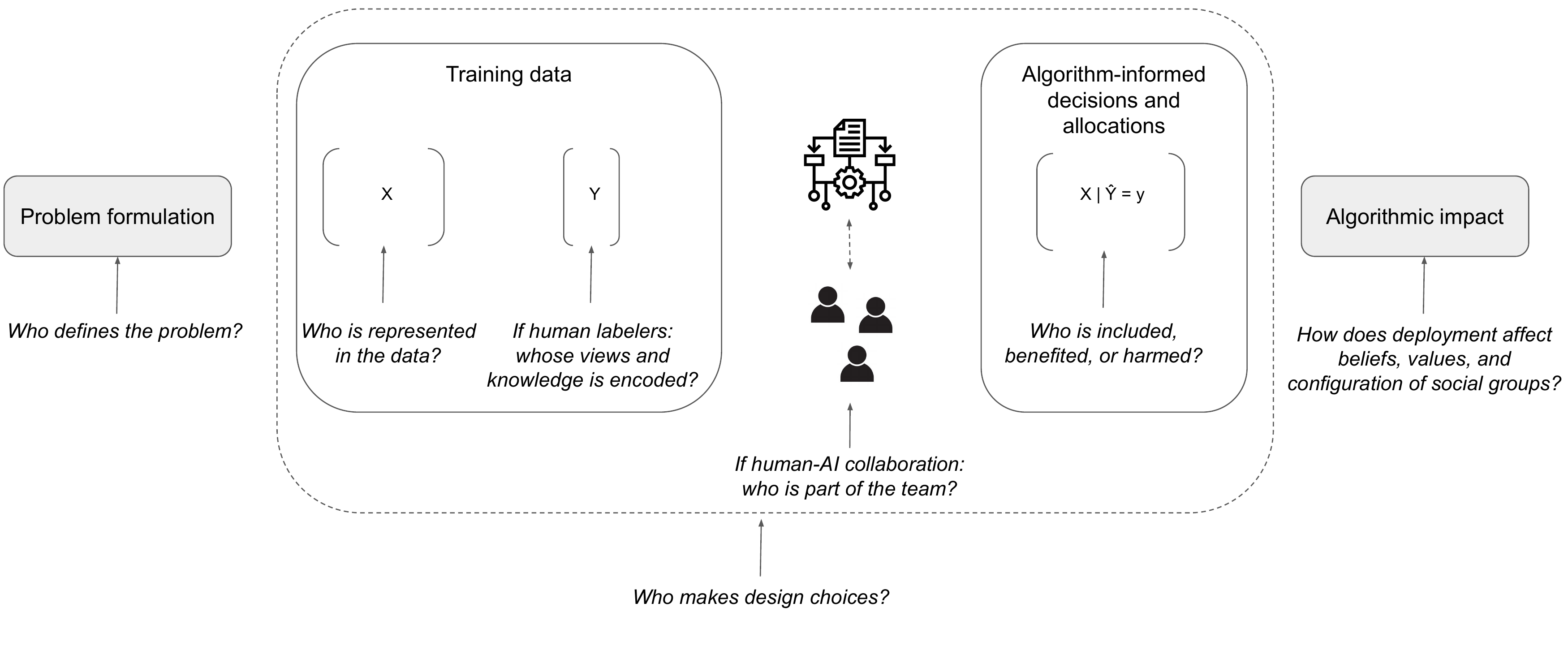}
    \caption{Stages of the lifecycle of a machine learning system at which different varieties of sociocultural diversity may be relevant.}
    \label{fig:lifecycle}
\end{figure}

The structure of this section is as follows: in each subsection, we focus on one stage in the design, development, and deployment pipeline. For each stage, we outline the nature and ramifications of the task, and delineate the set of diversity-related considerations that are pertinent to decisions and value judgments therein. \Cref{fig:lifecycle} provides an overview of these stages, and identifies diversity-related questions relevant to them.\footnote{We note that, while considering the possible existence of labels $Y$ centers supervised learning tasks, most of the discussion is relevant beyond supervised learning.} We explore potential ethical and political rationales that can support the use of different diversity concepts, and attendant measures, in particular settings. Moreover, when these diversity considerations pertain to (teams of) decision-making agents, we examine relevant cognitive and information elaboration-related factors that can influence their performance as socially situated agents, and identify potential ways in which increasing team diversity can benefit epistemic and ethical performance. Throughout we also highlight the strengths and limitations of existing work as well as promising avenues for future research. We postpone the discussion of broader issues to \Cref{sec:discussion}. 

\subsection{Problem formulation} 
Prior to the design and development of a ML system, its overarching goal and main function must be defined. This is frequently done by a team, or even an organization, that is distinct from the design and development team. 
For example, state governments tend to outsource the development of risk assessment models for recidivism prediction to external firms \cite{angwin2016machine}, while making in-house choices concerning the goal of the predictive model and the nature of its use. The task facing problem formulation teams includes defining a system's overarching goal, translating this goal into a prediction problem, defining the decision space available for prediction-informed decisions, and anticipating the potential impact of these decisions---possibly in concert with other interventions---on populations that will be affected by them \cite{mitchell2018prediction}. 
Algorithms that solve seemingly similar prediction problems can be embedded within entirely different problem formulations. For example, an algorithm for loan default prediction can be embedded in a broader system geared towards reducing the likelihood of default by providing pro-active academic support \cite{herr2005predicting}. 
Yet, a functionally similar algorithm can also be embedded as part of a system that aims to deter loan default ``with social stigma and shaming'' by means of prediction-guided public replies on borrowers’ social network \cite{ge2017}.

Accordingly, problem formulation is a consequential, complex process that consists of multiple, inter-dependent steps. Each step in the process may demand distinct types of domain knowledge as well as substantive value judgments, and it may be subject to potential disagreements \citep{coyle2020explaining}. Neglecting or underappreciating considerations of diversity can, therefore, result in problem formulation groups that advance the interests of only a small subset of stakeholders, exhibit significant overlap in their blind spots---ethical or otherwise---or preclude possibilities for effective participation among team members. 

In many high-stake domains, such as in the allocation of public resources, problem formulation often involves deliberative mini-publics and town halls \citep{pa_bulletin}. Details of the task and the social context of operation can inform the choice of relevant diversity concepts. For example, a representative notion of diversity may be justified in the context of a representative democracy. On the other hand, normic notions of diversity may be relevant if the proposed technology imposes a disproportionate risk on a community that has been historically disadvantaged by, or excluded from, key societal decisions. For example, in deliberations about the potential deployment of recidivism prediction algorithms as part of the US criminal justice system, considerations of social justice and anti-oppression can require that Black and Latinx, and Indigenous communities are represented in a greater proportion than what corresponds to their presence in the country’s population, precisely because these communities have been disproportionately harmed by the criminal justice system.\footnote{The motivations underlying a representative or a normic notion of diversity may also include a desire for legitimacy. Ensuring the participation of a group that has high levels of mistrust in the government may not only shape the resulting system, but it may also increase this community's perception of an algorithmic tool as trustworthy. This may have both positive and negative consequences, depending on whether the trust is warranted or the community engagement is primarily performative.} 

These considerations are crucial, even when problem formulation does not involve community engagement. In current discourse on diversity in responsible or ethical AI, the term is used interchangeably with ``balance'', ``representativeness'', and ``inclusion'' (particularly of marginalized communities) \citep{leavy2018gender, jobin2019global}. Yet, as noted in \Cref{sec:concepts}, these notions correspond to diversity concepts that are quite different in their meaning and appropriate operationalization. For example, as discussed in relation to standpoint theory, when researchers are concerned about including the viewpoints and insights of historically marginalized groups---gained through lived experiences and histories of advocacy---taking diversity seriously demands a normic conception (or some hybrid thereof). Here, a lack of clarity can lead to a mismatch between motivation and appropriate implementation. 

The variety of values, perspectives, and expertise involved in the process of problem formulation, in conjunction with the social setting of interactions, makes team diversity critical for this stage of ML construction. Some concrete examples will help to illustrate this point. Consider, first, that part of problem formulation involves identifying those aspects of current practices that are \textit{not} amenable to prediction-based treatment. For instance, considering algorithms used for advising students’ curriculum choices, \citet{aizenberg2020designing} note that while some aspects of this task can be specified in terms suitable for a prediction problem, the activity also involves ``practice-oriented aspects'' that are not easily of formalizable in this way. These aspects include ``building an informal and socially interactive companionship between the student and a human counselor, who can help the student elicit interests that are outside of her/his initial comfort zone and provide emotional support in moments of anxiety'' \citep[p. 5]{aizenberg2020designing}. This component may be especially important for students from certain marginalized backgrounds, such as first generation students. A problem formulation team that includes and effectively draws upon the knowledge and perspectives of relevant stakeholders (e.g., experienced counselors and student advocates) is likely to be better situated for identifying these practice-oriented aspects, and so the appropriate domain and mode of operation for ML systems. Yet, such considerations tend to go unnoticed in many sensitive applications, such as the increasing use of predictive algorithms for determining (the extent of) individuals' need for medical, unemployment, and disability benefits~\citep{disability-report}.

Even when some portion of the task appears amenable to prediction-based treatment, the particular formulation of the prediction problem requires careful anticipation and evaluation of its likely impact on different sections of society. This issue has become painfully salient in a recent example of bias in healthcare algorithms, where using health costs as a proxy for health needs resulted in systematic underestimation of the needs of Black patients, due to racial disparities in access to care and spending~\cite{obermeyer2019dissecting}. As previous research has highlighted, individuals' ability to anticipate and evaluate relevant hypothetical scenarios is likely to be correlated with their sociocultural identities and assumed social roles~\citep{fazelpour2020norms, catellani2021expert}. In participatory and value-sensitive design approaches, awareness of this type of correlations have traditionally led to an emphasis on team diversity in prototype and foresight studies~\citep{friedman2019value}. 
More recently, increased recognition of diversity's importance in these contexts has also led to proposals for the adoption of participatory and collaborative techniques in ML problem formulation~\citep{martin2020extending}. Ultimately, the appropriate use of these techniques will depend upon attention to the relevant concepts, measures, and pathways.

The examples above pertain to the potential benefits that are due to an association between sociocultural identities and the cognitive repertoires of team members. Information elaboration-based considerations are also relevant, since the deliberative interaction among teams takes place in institutional and societal settings subject to various group dynamics. A particular challenge here is ensuring the \textit{effective} participation of different team members throughout the deliberative process \citep{aizenberg2020designing}. Addressing this challenge requires adopting strategies that enable individuals, who might otherwise remain silent or feel excluded, to voice their opinions and concerns \citep{van2014participatory}. Diversity's influence on performance via information elaboration pathways thus gains immediate relevance in this respect. More studies are needed to explore these potential benefits in the context of ML problem formulation teams. 

\subsection{Design and development}
Once a problem has been formulated, a series of design choices ought to be made. These choices concern the data used for training the model, the performance metrics optimized, and the models considered for prediction. Similar to the problem formulation phase, these decisions may be distributed across multiple teams or may all be the responsibility of one team. 
Unlike many of the decisions involved in problem formulation, however, this stage requires significant machine learning expertise. 
It is thus useful to be specific about the potential pathways and relevant effect modifiers that can govern diversity's beneficial impact on performance (broadly understood) in the design and development stage.

Both cognitive and information elaboration pathways discussed in Section~\ref{sec:benefits} may impact algorithmic design. 
For instance, designers' identity and lived experiences may inform their ability to anticipate modes of failure. If this is so, then we may expect marginalized standpoints to better identify the inimical impacts of certain choices, such as the training data and the metric used to assess performance. 
But we should not expect diversity to be a panacea, and the pathways and effect modifiers that mediate diversity's impact should be attended to. 

Consider, for instance, a recent study of whether fairness properties of algorithms is impacted by the demographic attributes of programmers who train them~\citep{cowgill2020biased}. 
The authors ``found no evidence that female, minority and low-IAT [Implicit Association Test] engineers exhibit lower bias or discrimination in their code''~\citep{cowgill2020biased}. Instead, biased predictions were found to be ``mostly caused by biased training data''~\citep{cowgill2020biased}. Taken at face value, the findings seem to suggest that we should not expect sociocultural diversity of design teams to result in any epistemic benefits. 
It is crucial, however, to interpret the findings in relation to the setup of the experiments. Importantly, engineers worked alone within an experimental setting that pre-defined the problem formulation and central choices of the design, including (1)~the predictive task: predicting math performance using biographical features, (2)~the evaluation metric: mean squared error, and (3)~the covariates used, which had their names obfuscated. 
This setup is representative of many real world settings, in which engineers are tasked with solving a problem without being involved in many of the formulation and design choices. Nonetheless, the setup significantly restricts the diversity-related inferences that can be drawn from the findings. Without any group interaction, for example, information elaboration pathways are absent. With respect to cognitive pathways, moreover, the highly regimented nature of the task is likely to attenuate any potential benefits of diversity. 
Given this setup, the findings are thus fully consistent with literature showing that the effects of diversity are dependent on the nature and complexity of the task~\citep{van2012defying}. Appropriately interpreted, then, the findings highlight the importance of being specific about the relevant pathways 
when designing diversity initiatives for improving ML products, but should not be misinterpreted to mean that such potentially beneficial impacts do not exist.

Currently, there is a growing number of voices and initiatives calling for a more diverse workforce in AI, in part motivated by the thesis that this would lead to less biased, more ethical algorithms~\citep{jobin2019global}. The lack of clarity in many of these calls regarding pathways through which these benefits may be realized is harmful: it lends itself to performative diversity initiatives that have no tangible impact on the technology built, and can also lead to misleading inferences from findings about diversity's epistemic impacts. Clarity about pathways of diversity's benefits is thus a first step---both in research and practice---towards designing impactful diversity interventions.

\subsection{Training data}
\label{subsec:ml_training}
Data is a fundamental element of machine learning pipelines. In this section, we discuss diversity considerations related to the population who is represented in the data, as well as the collective in charge of providing labels for it, starting with the latter case. 

\paragraph{Labelers.} The labels used to train machine learning models are often generated by humans. This includes crowdsourcing~\citep{Gray.2019, doan2011crowdsourcing}, eliciting information from expert panels~\citep{gulshan2016development, mckinney2020international}, and learning from historical observational data of human decisions~\citep{de2021leveraging}. While target labels are often termed ``ground truth", labels derived from human assessments are subject to noise~\citep{callison2009fast}, expert disagreement~\citep{Adamson.2019}, and bias~\citep{Dressel.2018}. Accordingly, especially when labels are based on human appraisals, it is crucial to ask whose views, knowledge, and biases are being encoded. Attending to sociocultural diversity is particularly important--yet understudied--in tasks where the sociocultural identity of experts can be expected to impact their appraisals, or in settings where identity can instigate group influences that impact the quality of information elaboration of panels or groups in charge of labeling. 

Consider, then, diversity's potential benefits via the cognitive pathways. The requisite association between labelers' identities and their task-relevant knowledge and perspectives might emerge for a variety of reasons. 
In healthcare, for example, it has been shown that women, Black, and Latinx physicians are more likely to serve communities that have been historically underserved by the healthcare system  \citep{moy1995physician, komaromy1996role, cantor1996physician}. In the mid-90's, this was used as evidence to argue that dismantling affirmative action in medicine could imperil access to care for Black, Latinx, and low-income communities~\citep{komaromy1996role, cantor1996physician}.
Similarly, it is not difficult to see how drawing on this variety of expertise could enable health algorithms to better serve the needs of diverse patient populations. 

Diversity's benefits via cognitive pathways can also be relevant in crowdsourcing \citep{duan2020does}. Researchers studying data labeling for hate speech detection have found that disparities in performance across African American and non-African American English vary across annotators~\citep{keswani2021towards}. It is plausible that labelers' identity could in part explain the gap in their performance across different dialects. Thus, being attentive to the diversity of a group of labelers could help address the documented racial bias in automated hate speech detection~\citep{sap2019risk}.\footnote{Diversity of labelers has also been considered on ethical grounds as a path to ``rehumanize crowdsourcing" by emphasizing the relevance sociocultural identity and compensation in the annotator selection process~\citep{barbosa2019rehumanized}.} 

As before, the appropriate conception of diversity depends on the context and the task. A representative concept of diversity could be epistemically motivated, for example, if it is believed that each individual is best positioned to annotate instances written in their own dialect. 
Consider the task of hate speech annotation in a Spanish corpus, where terms vary widely across countries and regions~\citep{rodriguez2018dialectones}. Ideally, one may want to match annotators who are familiar with the local dialect to be tasked with annotating instances written in said dialect. In other contexts, however, a different diversity concept might be more appropriate. Consider that hate speech annotation introduces a second dimension: the legacy and pervasiveness of racist and other discriminatory terms, whose derogatory and demeaning nature tends to go unnoticed by the majority in everyday and even professional discourse~\citep{houghton2018blacklists, aspinall2005language}. Here, those who are more likely to be victims of hate speech may be better positioned to identify it.\footnote{This presents a tension with the mental health toll of content moderation and annotation, which may be heightened for those who are targets of the attacks.}
Such cases can be seen as an instance of standpoint theorists' epistemic advantage thesis; cases in which those occupying historically marginalized standpoints are better positioned to evaluate not only their own dialects, but also identify and address the blind spots of the dominant culture. In these contexts, therefore, a normic concept of diversity might be more appropriate. 

Finally, when considering expert panels involved in data labeling, in addition to task-relevant cognitive pathways, one must consider information elaboration pathways. Here, socioculturally heterogeneous groups of labelers may benefit from better information sharing, reduced normative conformity pressures, and enhanced critical assessment. Nonetheless, as mentioned in \Cref{sec:benefits}, translating this research into practice also requires attending to the unique challenges and performance trade-offs associated with diversity (see, e.g., \cite{duan2020does}). An important first step towards understanding the impact of the group diversity on the resulting labels is to standardize practices of reporting demographic information for panels of labelers. 

An important open question is how to account for the role that experts may play in mediating observed outcomes. Research has found that Black men seen by Black doctors are more likely to agree to preventive services that can reduce cardiovascular mortality~\citep{alsan2019does}. Similarly, the increase in women police officers has been linked to improved rates of crime reporting and reduced domestic violence~\citep{miller2019female}. Thus, sociocultural diversity may be relevant to target labels even when these do not directly correspond to human assessments, but are mediated by them.  

\paragraph{Data instances.} Up to this point, we have discussed diversity in relation to teams and collectives whose views and knowledge shape the ML technology. An additional set of diversity considerations relates to items (e.g., individuals, regions, ...) represented in the data used to train and validate predictive algorithms. Given the growing recognition of the implications of a dataset's diversity for the resulting algorithm's performance and fairness-related properties, diversity as a design desideratum in this stage of the pipeline has perhaps received the most attention within the machine learning literature~\citep{celis2016fair, shankar2017no, buolamwini2018gender, atwood2020inclusive, asudeh2019assessing}. 
\footnote{In this stage of the pipeline, the previously discussed epistemic rationales are less relevant. In particular, the cognitive and informational pathways that mediate the impact of diversity are no longer present, since the set whose diversity is under consideration is not engaged in knowledge-seeking or decision-making tasks. Instead, the pathway that mediates the impact is the algorithm itself, together with any pre-processing and post-processing steps.} 
While existing work in this space has approached diversity using different measures and with different goals in mind, there is frequently an implicit assumption that diversity is a monolithic concept. However, the question ``is this dataset diverse?'' does not have a single answer, as the answer will depend on the concept under consideration---a choice that is distinct from and ought to precede the choice of a diversity measure. 

Being explicit about the underlying concept of diversity (and its rationales) is important when exploring potential tensions between diversity and other design desiderata. 
Consider curating a dataset with the aim of maximizing (i) sociocultural diversity of individuals represented therein, and (ii) coverage over the entire feature space to improve generalizability. While some works have suggested a trade-off between these two~\citep{celis2016fair}, other works claim that they are in alignment~\citep{asudeh2019assessing}. This apparent contradiction can be resolved by acknowledging the distinct concepts of diversity implicit in these works. In particular, representativeness of the entire feature space is not in conflict with diversity in a \textit{representative} sense. That is, while the representative concept of (sociocultural) diversity considers a subset of the features, it nonetheless has the same goal and can be estimated using the same families of measures. The tension arises, however, when considering \textit{egalitarian} concepts of diversity, which will be in conflict with the aim of representativeness whenever the distribution across groups in the reference population is not uniform.

A monolithic understanding of the notion of diversity can also result in misinterpreting potential tensions between diversity and fairness-related desiderata. 
Suppose, for example, that an ethnic group makes up a small percentage of a relevant population. 
Here, assuming a representative diversity concept--as is often implied in work on diversity of training data~\citep{shankar2017no, buolamwini2018gender}---would mean that this group should be represented in the training data in the same proportion as in the population. Yet, this representatively diverse sampling could lead to disparities in performance, and thus be undesirable from an algorithmic fairness perspective. 
This does not mean, however, that there is an inherent trade-off between fairness and diversity per se. Rather, such a trade-off between representative diversity and fairness is relevant only to the extent that there is indeed an epistemic or ethical rationale for representatives diversity as the contextually appropriate conception of diversity, which may not always be the case. Acknowledging the variety of diversity concepts and clearly articulating their underlying justifications can thus help avoid confusions. 

\subsection{Human-AI teams}
In many use cases, the function of predictive algorithms is to offer informational support to human decision-makers who---either as individuals or teams---make the ultimate decisions. In such settings, diversity-related considerations arise in relation to the characteristics of the human users, the AI tool, and the interaction between the two. 

In Section~\ref{subsec:ml_training}, we discussed potential associations between labelers' sociocultural identity and their assessments. This association can also be relevant to how individuals integrate AI recommendations into their decisions. Research has shown two mechanisms through which this may occur: social context may result in differential rates of adherence to algorithmic recommendations~\citep{albright2019if}, and demographic attributes may be associated with how decision makers prioritize and integrate different sources of information~\citep{mallari2020look, peng2019you}. In the context of judges' adherence to risk assessment instruments,~\citet{albright2019if} shows that judges' likelihood of overriding algorithmic recommendations in favor of harsher bond conditions vary across geographic locations. What is more, this variation is shown to exacerbate disparities across racial groups, since cases of override are more common in counties with larger Black populations.\footnote{Note that this is different from the effect of the defendant’s race, which is also studied by~\citet{albright2019if}.} Also studying recidivism prediction--but this time in the context of an Mturk study--\citet{mallari2020look} show that decision-makers' self-identified gender was a significant factor in recidivism predictions, and that this interacted with demographic attributes of individuals subject to those decisions. 
The gender of decision-makers was also shown to impact algorithm-informed decisions (and associated biases) in an MTurk study assessing the use of algorithmic hiring tools~\citep{peng2019you}. Thus, attending to diversity considerations in relation to those relying on ML recommendations is central to the study of algorithmic adoption, both to ensure the validity and generalizability of empirical studies and the effective and responsible deployment of predictive algorithms.

Additional diversity considerations become salient once we recognize the \textit{human-AI teams} as the the primary unit whose performance is of interest. That is, instead of assessing the behavior and performance of algorithms or human users in isolation from each other, we might attend to the human-AI team as a unit and consider how the constituents of the unit \textit{complement} one another towards achieving a shared task. 
As discussed in \Cref{sec:benefits}, such complementarity between the task-relevant attributes of team members is central to how cognitive diversity can enable a productive division of labor and so improve team performance on complex tasks. The same logic applies to the case of human-AI teams. For example, ~\citet{madras2018predict} show that substantial performance gains can be made when predictive algorithms are optimized to prioritize correct predictions on instances that are difficult for human experts, while deferring instances that are challenging for the algorithm to human experts (see also~\citep{wilder2020learning}). Importantly, researchers have shown that these gains can be amplified by attending to the heterogeneous skills of experts, when choosing \textit{which} expert to defer to~\citep{gao2021human}. In line with the discussion in \Cref{sec:benefits}, whether and when such performance gains could be achieved generally depends on the nature and difficulty of the task as well as the type of errors made by humans and algorithms (see~\citet{tan2018investigating}). An interesting and understudied issue here pertains to the development of methods that explicitly consider and benefit from the demographic diversity of human experts.

\subsection{Algorithm-informed decisions and impact} 

Whether an algorithm is making autonomous decisions, or assisting human decision-makers, the diversity of its output set is also important. This topic has been considered in works on ranking and information retrieval~\citep{singh2018fairness, karako2018using}, and more broadly on subset selection \citep{mitchell2020diversity}. In the context of image retrieval, one may care about the diversity of a set of images shown; in targeted advertisement, the diversity of those who will be shown a job ad matters; and if ML is used to pre-select individuals who will be hired or admitted into a program, the diversity of the selected set is of key relevance. As before, the relevant concept of diversity will depend on the task that the algorithm is solving. 

Representative notions are frequently evoked when discussing diversity of algorithmic output, and are especially attractive in the context of information retrieval systems~\citep{singh2018fairness}, where the task is often construed as that of ``mirroring'' the world. That is, if a search engine is believed to be a descriptive tool, an argument for a representative conception of diversity can follow naturally. This argument may falter, however, if demographic information is deemed conceptually irrelevant to a query, in which case an egalitarian concept may be justified. Similar to a dictionary, which does not define ``surgeon" to be a man, irrespective of the frequency of different genders in this occupation, one may expect a search engine to provide query-relevant results that are unaffected by such contingent frequencies.  

Importantly, the societal impact of algorithms goes beyond those directly subjected to algorithmic predictions. For instance, information retrieval and search engines are not purely descriptive tools; instead, they hold an active role in shaping beliefs and behaviors. Considering this may significantly impact the concepts of diversity that are appropriate. Thus, normic (and hybrid) notions become more relevant once we consider the active role of such systems in shaping societal beliefs. 
Indeed, in one of the most in-depth discussions of diversity in algorithmic outputs and unlike most other approaches, ~\citet{mitchell2020diversity} defend a normic notion of diversity. In so doing, they introduce a family of metrics that facilitate the quantification of diversity in algorithmic subset selection from a normic perspective, grounded in considerations of social power differentials.

Systems of incentives introduced by the deployment of an algorithm may also have significant impacts on diversity. For instance, algorithmic unfairness may disincentivize investment from individuals who invest rationally~\citep{Liu.2020}. As an example, if the use of standardized tests in college admissions disadvantages a group, this group has less incentives to invest in standardized testing, which would impact the diversity of both the pool of college students and the pool of applicants. Furthermore, in the presence of multiple players--such as multiple companies making hiring decisions--research has shown that a partial compliance to (supposedly) fairness-enhancing interventions can result in segregation~\citep{Dai.2021}. Thus, considering dynamics of deployment is crucial to understand the impact of algorithms on the diversity of different groups in society.

\section{Discussion} \label{sec:discussion}
\subsection{The meaning of measures}
Throughout, we have emphasized the importance of achieving clarity about the specific diversity concept that underpins a proposed measure in a given context. It is also important to note that when the same mathematical formula is used for measuring different concepts, this should not be misinterpreted as an equivalence between these concepts. Consider, for example, the use of demographic parity in the context of ethical evaluation of algorithms. Demographic parity is both a popular measure for representative notions of diversity and a measure of disparate impact (or indirect discrimination). To be sure, when the appropriate reference populations are different, the mathematical formula can output different numbers in the case of diversity and discrimination  (e.g., if evaluating representative diversity pertains to a comparison with general population, but assessing disparate impact requires considering an applicant pool) \citep{drosou2017diversity}. 
Even when the reference population is the same, however, the interpretation of demographic parity depends on which concept---representative diversity or disparate impact---it is meant to measure. 
In the case of representative diversity, once we have a well-justified reference population, the extent to which a collective matches that reference population allows us to immediately make claims about its (representative) diversity. In contrast, in the case of disparate impact we typically cannot make the same immediate inference. For instance, the 80 percent rule \citep{Barocas.2016} considers a sufficiently large deviation from demographic parity to be a \emph{presumption} of disparate impact. Whether this is indicative of discrimination, however, depends on properties of the process, which cannot be determined by this metric alone. In other instances, there might be a closer connection between measures of diversity and fairness, as may be the case in the selection of participants for a deliberative mini-public in the context of a representative democracy. Importantly, though, this close connection is a result of underlying conceptual similarity, insofar as fairness of the process is grounded in equal representation. Maintaining clarity over conceptual differences can prevent erroneous conclusions on the basis of metrics alone.

\subsection{Navigating value tensions}
The discussion so far has eluded a key difficulty. Consider, for instance, that whether, and to what extent, team diversity can result in epistemic benefits critically depends on how we measure performance. But what if individuals---perhaps in ways that correlate with sociocultural identity---cannot agree on a performance measure? In the context of fair ML, these potential disagreements can arise when individuals disagree about which fairness measure is appropriate for evaluating their performance as a team in relation to the aim of reducing bias. This can become particularly challenging when the disagreed upon measures cannot be jointly satisfied in context \citep{chouldechova2017fair}. Similarly, while diverse groups are more likely to come up with better solutions (in variety and quality), they may also take more time to reach consensus \citep{fazelpour2021diversity} or never reach one at all (e.g., due to polarization) \citep{muldoon2018paradox}. Yet, these additional considerations---whether and when a consensus is reached---are clearly crucial to social and institutional planners working under time and resource constraints. A more fundamental framing of the issue is this: researchers have offered different ethical \textit{and} epistemic rationales in support of distinct diversity concepts. While in many contexts these rationales work in tandem, it is also possible that they come apart. What should we do when these rationales are in conflict? We have no context-free solution to these value tensions. Nonetheless, two general points are worth emphasizing.

First, such value tensions are widespread in the context of deliberative and collaborative groups. A useful way to proceed, therefore, is to draw on techniques developed in previous research. For example, researchers in value-sensitive design have long had to contend with these value tensions, and so have developed a range of techniques for addressing them \citep{friedman2019value}. In many contexts, for instance, it might be useful to shift the focus of discussion from underlying values to proposed courses of action. This can be helpful because actions are often over-determined by values---that is, divergent value systems can nonetheless converge on the same action as the appropriate one in context (even if they disagree about \textit{why} this is so) (see also \citep{hansson2017theories}). Similarly, recent works on diversity in political philosophy have sought to devise bargaining techniques for addressing disagreements among individuals with fundamentally different perspectives (e.g., about the very conception of the task) in ways that are nonetheless acceptable to those involved \citep{muldoon2016social}. 

Second, in thinking about these value tensions, we caution against the trap of myopia. Viewed from a static or short-term perspective, such value tensions might appear inescapable. Things may look different, however, when we broaden our purview. This is familiar from works that examine the situated dynamics of algorithmic decision-making. When we move beyond the static setting of one-shot classifications to consider strategic plans consisting of multiple interventions over time, for instance, we can satisfy fairness measures that appear irreconcilable when considered statically \citep{hu2018short,fazelpour2020algorithmic}. Similar points have been raised in political philosophy in relation to justice and diversity \citep{anderson2010imperative}. As noted in \Cref{sec:benefits}, moreover, enabling conditions for realizing diversity's potential epistemic benefits embody values of respect, non-domination, participation, and inclusion. Of course, given our current world, instituting and supporting these values takes time and sustained commitment. But this also highlights that the tension between ethical and epistemic aims is often not just that; it is also a tension between short- and long-term epistemic considerations. Ultimately, how tensions are resolved and what trade-offs are struck depends on our value judgments. We hope that the foregoing offers some tools for making these judgments in an informed way, while elucidating (some of) the normative commitments involved therein. 
\section{Conclusion} \label{sec:conclusion}

Awareness of the importance of diversity at different stages of the ML lifecycle has increasingly featured in discussions about AI ethics. Meaningful conversations, studies and interventions hinge on our ability to clearly define and articulate diversity and its implications. Researchers in philosophy, psychology, and social and organizational sciences have long worked to understand diversity's varied meanings and specify pathways by which diversity can be functionally beneficial to groups. This paper draws from this literature to bring attention to how current research in and on ML is hampered by a lack of clarity about the underlying concepts of diversity and precise pathways of its beneficial consequences. We have first introduced a taxonomy of concepts of diversity from feminist philosophy, purpose-built for understanding \textit{sociocultural} diversity, followed by an overview of multi-disciplinary findings concerning benefits of diversity and the pathways through which these may be realized. Building on this, we have brought this conceptualization into the lifecycle of ML systems, providing an overview of the diversity-related questions that feature at each stage. Here, we have shown that clarity over pathways and concepts can resolve seemingly contradictory findings and impossibility results in existing work, and enable better system design. 

By providing a detailed characterization of the various contexts in which diversity can be relevant throughout the ML lifecycle, this paper makes a case for diversity as a design desideratum of teams, data and models. By translating cross-disciplinary literature and bringing it to bear on the study of sociotechnical ML systems, it provides conceptual tools to further advance research and practice grounded in a coherent understanding of sociocultural diversity. 

\bibliography{refs}

\end{document}